\documentclass[aps,rmp,twocolumn,twoside,showpacs,showkeys,preprintnumbers,superscriptaddress,amsfonts,amsmath,amssymb,floatfix,a4paper]{revtex4-1}

\usepackage[dvips]{graphicx} 
\usepackage{dcolumn} 
\usepackage{bm}
\usepackage{amsmath}
\usepackage{amssymb}
\usepackage{color}

\newcommand{\Rh}{\ensuremath{a}}

\newcommand{\Sec}[1]{Sec.~\ref{sec:#1}}

\newcommand{\Eq}[1]{Eq.~(\ref{eq:#1})}

\newcommand{\Fig}[1]{Fig.~\ref{fig:#1}}

\newcommand{\ie}{\emph{i.e.,}~}

\newcommand{\cf}{\emph{cf.}~}

\newcommand{\rme}{\ensuremath{\mathrm{e}}}

\def \H {\ensuremath{\hbox{H}_2}}
\def \He {\hbox{He}}

\begin{document}

\title{The journey of hydrogen to quantized vortex cores}
\author{Gregory P. Bewley}
\affiliation{Max-Planck-Institut f\"ur Dynamik und Selbstorganisation (MPIDS), 37077 G\"ottingen, Germany}

\author{J\"urgen Vollmer}
\affiliation{Max-Planck-Institut f\"ur Dynamik und Selbstorganisation (MPIDS), 37077 G\"ottingen, Germany}
\affiliation{Fakult\"at f\"ur Physik, Georg-August-Universit\"at G\"ottingen, 37077 G\"ottingen, Germany}

\date{\today}


\begin{abstract}

Nanoscale hydrogen particles in superfluid helium track the motions of quantized vortices.  
This provides a way to visualize turbulence in the superfluid.  
Here, we trace the evolution of the hydrogen from a gas to frozen particles migrating 
toward the cores of quantized vortices.  
Not only are the intervening processes interesting in their own right, 
but understanding them better leads to more revealing experiments.  

\end{abstract}


 \maketitle

\section{Introduction}

What makes liquid helium particularly useful for studies of fluid dynamics and turbulence 
is its small viscosity \cite[][]{skrbek:2004, niemela:2006}.  
Another interesting aspect of liquid helium is that in the superfluid phase circulation is quantized, 
and turbulence arises in it through the formation and interaction of quantized vortices \cite[][]{feynman:1955}.  
The superfluid is present in liquid 
helium-4 at temperatures between 0\,K and T$_\lambda$, 
where T$_\lambda \approx$~2.17\,K when the liquid and vapor coexist \cite[][]{donnelly:1991}.  

The strides made in understanding classical turbulence have 
not been matched in quantum turbulence.  
Challenges lie in both the theoretical description of the fluid motions and in the 
acquisition and interpretation of experimental data \cite[][]{vinen:2002}.  
In an effort to surmount the latter hurdle, one of the authors (GB) 
developed a novel hydrogen-particle technique for visualizing the internal motions of 
superfluid helium.  
The technique involves 
the generation of the particles {\it in situ}, and sequential recording 
of their positions \cite[][]{bewley:2009b}.  

What experimenters observe is the following.  
Nanoscale hydrogen particles appear in liquid helium close to its boiling point, 
after injection of a mixture of hydrogen and helium gas through a small tube.  
When the temperature is lower than T$_\lambda$, 
the particles gravitate toward quantized vortex cores.  
The entrapment does not occur when the temperature 
is close to T$_\lambda$, but only at significantly lower temperatures \cite[][]{bewley:2006b}.  

The motions of quantized vortices become visible 
because particles become trapped on their cores \cite[][]{parks:1966}.  
Once they are visible, it is possible to observe the vortex lattice in 
rotating helium \cite[][]{bewley:2006}, vortex reconnection 
\cite[][]{bewley:2008b}, and vortex ring decay \cite[][]{bewley:2009a}.  
An increasing number of laboratories around the world are employing similar techniques.  
They do so for technological reasons, 
such as for the generation of nano-wires \cite[][]{gordon:2012}, 
but above all to study the fluid dynamics of helium.  

Since the technique is employed in many ways, it makes sense to better understand it.  
The self-organization of hydrogen onto quantized vortices 
involves non-equilibrium processes occurring on many length scales.  
These start from the microscopic physics of demixing and freezing that produce the particles, 
to the mixing of the particles into the liquid 
and the migration of the particles toward the cores of vortices, 
to the macroscopic turbulent motions of the helium itself.  
This paper proposes descriptions for each process.  
These should be understood to be conjectural, 
as we have not tested the ideas experimentally.  

Specifically, we set out to explain how the hydrogen particles form, 
how they end up in the liquid rather than the gas phase of helium, 
and why vortices do not entrap them when the temperature is close to T$_\lambda$.  
In \Sec{setup}, we describe the experimental apparatus.  
In \Sec{history}, we survey briefly recent work 
that exploits particle-imaging to study superfluid helium dynamics.  
In \Sec{partgen}, we describe the phase transitions that produce hydrogen particles, 
and in \Sec{partdep} the process dispersing the particles in the liquid helium.  
In \Sec{thermo}, we argue that thermodynamics does not limit the trapping 
of particles unless the temperature is very close to T$_\lambda$, 
and in \Sec{dynamics} we argue that particle dynamics, 
rather than thermodynamics, limit the ability to visualize vortices.  
Finally, in \Sec{conc} we comment on these interpretations.

\section{Hydrogen particle generation revisited}

\subsection{Experimental setup}
\label{sec:setup}

\citet{lamantia:2012} and \citet{bewley:2009b} 
describe the relevant experimental apparatus, 
which we briefly review here.  
Hydrogen particles are created from a room temperature mixture of 
hydrogen gas diluted with helium.  
The method works also with other 
gases, such as deuterium, in place of hydrogen.  
The hydrogen-helium mixture ratio controls the size of the particles, 
the size increasing with the ratio of hydrogen to helium.  
Mixture ratios of the order of 1\,\H\,:\,100\,\He\, by volume produce particles about 
1\,$\mu$m in diameter.  
The mixture is released into liquid helium 
through valves that control the injection rate and duration.  
For a 600\,mL volume of liquid 
helium in a cryostat, 20\,mg of the mixture injected over a period of 
5\,s yields a final particle volume fraction about $10^5$ times 
smaller than the total volume, which corresponds to a few million 
particles per cm$^3$.  
The gas mixture passes through a tube 
that opens somewhere beneath the free surface of the helium, 
so that the mixture bubbles through the liquid.

\subsection{Recent history}
\label{sec:history}

We cite a few papers to give an impression of the field's breadth.  
\citet{paoletti:2008} shows that the distribution 
of particle velocities in superfluid helium is non-Gaussian, in 
contrast to the near Gaussian statistics of particle motions in 
classical flows.  
\citet{chagovets:2011}
measures in a counterflow the heat flux required to dislodge hydrogen 
particles from vortices.  
\citet{jin:2011} describes some 
interesting behavior of dust particles in superfluid helium, and 
\citet{lamantia:2012} confirms that hydrogen particle 
velocities equal the normal fluid velocity at low heat fluxes.  
\citet{fuzier:2008} shows that superfluid helium can be 
used to segregate particles by size.  
\citet{guo:2009}
images electron bubbles and speculates that this technique could 
be used to track quantized vortices.  
Metastable helium molecules 
can also be tracked, an approach employed by 
\citet{guo:2010} to measure the velocity profile of superfluid 
helium in a channel.  
\citet{gordon:2012} discusses the 
role of vortices in the generation of metallic nanocrystals and 
nanowires in superfluid helium.  
Taken together, the collected works 
indicate that under some conditions, such as in weak 
counterflows, quantized vortices trap particles on their cores, while 
in other conditions, the particle-vortex interactions yield more 
complicated behavior.  
At low particle number densities and when the 
particles are trapped, the particle motions reveal that vortex 
reconnection is a self-similar process \cite[][]{bewley:2008b}, while at high number densities, 
networks are formed where crossing vortices stick rather 
than reconnect \cite[][]{bewley:2006}.  

The behavior of particles in superfluid helium and in particular the 
interaction between particles and quantized vortices have been the
subject of many numerical and theoretical investigations.  
Notable examples include 
 \citeauthor{poole:2005}'s (\citeyear{poole:2005}) discussion of the motions of particles 
in the context of the equations of particle motion.  
\citet{fujiyama:2007} demonstrates the pinning of quantized 
vortices by solid spheres, an interaction explored further by 
\citet{kivotides:2008}.  
\citet{barenghi:2009} and 
\citet{bewley:2009a}
consider the influence of bound particles on quantized vortex dynamics.  
This work inspired interpretations of experimental observations.  
It also argued that particles are trapped when 
the relative velocity between the normal fluid and the quantized vortex is not too large, 
and that the process of particle trapping by a 
quantized vortex curves the vortex core and causes Kelvin waves along the core.  
Once trapped, the particles increase the strength of the coupling between the 
quantized vortices and the normal fluid.

\section{Physical Mechanisms}
\label{sec:physmech}

No one has yet found a way to observe particle formation directly.  
In the following, we lay out our personal view of the underlying mechanisms, 
which involve physical problems worth studying in their own right.

\subsection{Particle generation}
\label{sec:partgen}

In a typical hydrogen particle generation experiment, gas is 
injected at a rate of $J = 4 \times 10^{-3}$\,g/s through a pipe with a 
diameter of about 0.15\,cm.  
Given the mass density, $\rho_g \approx 1.5 \times 10^{-3}$\,g/cm$^3$, 
this amounts to a flow velocity of 14\,cm/s.  
For a 1\,m tall cryostat, the gas takes about 7\,s to 
exit the tube into the liquid helium bath.  
As it descends through the injector into the cryostat, 
the mixture of hydrogen and helium gas has a higher temperature 
than the ambient helium bath.  
The generation of particles should therefore be understood as a problem of cooling 
predominantly helium gas bubbles that cool until hydrogen condenses.  
In such a setting, we expect that the initial hydrogen particles 
are created by homogeneous nucleation \cite[][]{doremusBook}.  
This process occurs in analogous binary fluids cooled into the miscibility gap 
\cite[][]{vollmer07PRL, LappRohloffVollmerHof2012}.  
At the low temperatures of helium, quantum tunneling may 
promote crystallization of the hydrogen \cite[][]{levi:2001}.  

Specifically, the hydrogen starts as a warm gas strongly diluted by helium.  
After injection, the gas mixture bubbles cool as they rise through the liquid helium.  
The melting and boiling temperatures of hydrogen at the relevant pressures 
(e.g. 14\,K and 22\,K at 1\,bar) 
exceed the boiling point of helium (4.2\,K at 1\,bar), so that the 
hydrogen condenses out of the mixture even while the helium fraction remains gaseous.  
Cooling proceeds even after particle nucleation.  
In this way, the condensation is similar to 
the formation and growth of rain droplets in orographic 
rain formation, where clouds cools slowly due to adiabatic expansion 
\cite[e.g.][]{shaw:2003}, to the formation of droplets in laboratory experiments where 
binary fluid mixtures are slowly pushed into a phase coexistence region by 
an applied temperature drift \cite[][]{vollmer07PRL, LappRohloffVollmerHof2012}, 
and to the formation of slow dust in Enceladus' plume from the condensation of water 
\cite[][]{schmidt:2008}.  
Understanding of phase separation in such settings is presently a 
subject of active research due to the importance of clouds in 
climate dynamics \cite[][]{bodenschatz:2010, Pierrehumbert2010}.  
Based on the studies of phase separation in the presence of a 
temperature drift, we expect a wide distribution of particle sizes 
\cite[e.g.][]{LappRohloffVollmerHof2012}.  
The distribution results from the 
interplay of growth and ripening \cite[][]{LifSly61, Wagner1961, Slezov2009}.

\begin{figure}
\includegraphics[width=0.45\textwidth]{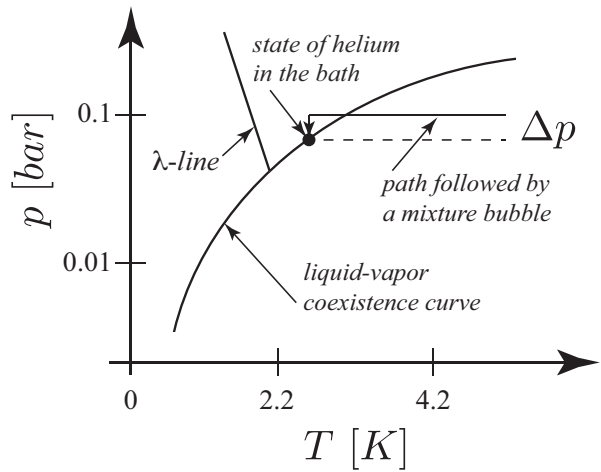}
\caption{Using a schematic phase diagram, we follow the path of a 
  helium bubble as it cools while rising through liquid helium.  
  When the influence of the hydrogen is neglected, 
  the liquid-vapor coexistence curve marks the boundary between 
  the gaseous (below) and liquid (above) phases of helium.  
  The $\lambda$-line marks the boundary between normal liquid helium 
  (to the right), and the liquid phase with superfluid properties (to the left).  
  The pressure difference between the bath and the bubble, 
  $\Delta p = 2 \sigma / R \approx 2 \times 10^{-6}$\,bar, 
  is due to surface tension \cite[][sec.~6.9]{AtkinsBook}, 
  and has some time dependence that we do not show.  
  Here, $\sigma$ is the surface tension and $R$ is the radius of the bubble.  
  When the bubble liquifies, hydrogen 
  particles suspended in the bubble are released.  
\label{fig:phasediagram}}
\end{figure}

\subsection{Deposition of the particles into the liquid}
\label{sec:partdep}

From the analogy with boiling water, one might expect that 
the bubbles would grow in size and then break through the liquid helium surface, 
dispersing the hydrogen particles in the gaseous phase of helium.  
However, a bubble is not stable in a liquid with a free surface, 
due to the high free energy associated with the surface tension.  
In experiments, bubbles grow smaller at a rate that is limited by 
how quickly the heat of the bubble can be transferred from it to the surrounding liquid.  
This occurs even if the system is on the liquid-vapor coexistence curve.  

The bubbles cool at approximately constant ambient pressure, 
as illustrated in \Fig{phasediagram}, 
due to the low density of liquid helium.  
One must consider six contributions to the heat of the bubble.  
These are due to the temperature differences between the surrounding helium 
and the hydrogen and helium in the bubble, 
the heat of condensation and of freezing of the hydrogen, 
the heat of condensation of the helium, 
and the heat released as the surface area of the bubble decreases.  

The key point is that each contribution to the free energy of the bubble 
constitutes an amount of heat that must be transferred to the surrounding liquid 
\cite[][]{ChaikinLubenskyBook}.  
The rate at which the bubble collapses is determined 
by the rate at which this 
heat can be carried away from the bubble by diffusion or convection.  
Initially, the mass of the bubble will increase due to this heat transfer, 
because some of the surrounding helium will evaporate into the bubble.  
Eventually, however, the temperature of the bubble is low enough that it condenses.  
Here, we consider each contribution in turn.  

The heat due to the initial warmth of the droplets is 
\[
\Delta Q_{\textrm{heat}}
\approx \frac{4 \pi}{3} \; R^3  \rho_g  C_p \: \Delta T
\gtrsim 4 \times 10^{-4}\: \textrm{J}
\]
where $\rho_g \approx 1.4\;$kg/m$^3$, 
$C_p \approx 1.2 \times 10^2\:$m$^2$/s$^2\,$K, and 
$\Delta T \gtrsim 10\;$K, are the gas density, its heat capacity and 
a conservative estimate of the initial temperature difference, respectively.  

The latent heat of the phase transitions include those from 
the fluidization and solidification of hydrogen, 
and from the fluidization of the helium in the bubble.  
They are 
$\Delta q^{H_2}_{\textrm{fluidize}} = 4.5\times 10^2\:$J/kg, 
$\Delta q^{H_2}_{\textrm{solidify}} = 6.0\times 10^4\:$J/kg, and 
$\Delta q^{He}_{\textrm{fluidize}}  = 2.1\times 10^4\:$J/kg, 
respectively.  
For the cooling of the bubbles they give rise to the contributions
\begin{eqnarray*}
\Delta Q &=&  \frac{4 \pi}{3} \; R^3  \rho \, \phi \; \Delta q
\\[2mm]
\Delta Q^{H_2}_{\textrm{fluidize}} & \approx &  1.3 \times 10^{-4}\:\textrm{J} \\[1mm]
\Delta Q^{H_2}_{\textrm{solidify}} & \approx &  8.9 \times 10^{-5}\:\textrm{J} \\[1mm]
\Delta Q^{He}_{\textrm{fluidize}}  & \approx &  1.2 \times 10^{-4}\:\textrm{J}
\end{eqnarray*}
where $\rho$, $\phi$ and $\Delta q$ denote the material constants for 
the respective materials and transitions; in particular, 
$\phi = 1/200$ is the mass fraction of hydrogen in the droplets.  
It is so small that $\phi \approx 1$ for the helium.  
The latent heat released upon going through these phase changes,
$\Delta Q_{\textrm{condensation}} 
= \Delta Q^{H_2}_{\textrm{fluidize}} 
+ \Delta Q^{H_2}_{\textrm{solidify}}
+ \Delta Q^{He}_{\textrm{fluidize}}$, 
accounts for about half of the total heat, 
$\Delta Q_{\textrm{tot}} 
= \Delta Q_{\textrm{condensation}} 
+ \Delta Q_{\textrm{heat}}$,
to be released from the droplets.  

The bubbles collapse when, upon rising, they dissipate the excess heat 
$\Delta Q_{\textrm{tot}}$ 
into the fluid helium bath before they reach the interface to the macroscopic gas phase.  
Note that the heat from the temperature mismatch of the bubbles 
exceeds the contribution from the surface energy 
by five orders of magnitude, such that the latter may be neglected.  

We estimate how long it takes for the bubble to collapse by 
comparing to the experimental observations of \citet{maris:1987} 
and to the numerical study of \citet{unsulu:2008}.  
Both studies treated pure hydrogen droplets in a liquid helium bath.  
\citet[see their Fig.~3]{maris:1987} observed that the temperature of droplets 
equilibriated in about $0.2\;$seconds.  
\citet{unsulu:2008} found that the droplets released their heat in a few seconds.  
We must bear in mind that the heat content of the hydrogen droplets 
is more than $25$ times that of helium bubbles, 
due to the very high heat of solidification of the hydrogen.  
Our mixtures of He and H$_2$ will therefore cool faster.  

How long does it take for the bubbles to reach the free surface in the cryostat, 
setting aside any changes in size?  
Due to the low viscosity of liquid helium, 
the bubbles generate a turbulent wake upon rising.  
The resulting rising velocity is approximately 
\[
U  
\approx  \left( 2 g R \; \frac{\rho_g}{\rho_f} \right)^{1/2}
= 1.4 \; \frac{\textrm{cm}}{\textrm{s}} \, .
\]
One can check that the assumption of a turbulent wake is justified.  
Given the density, $\rho_f = 140\;$kg/m$^3$, and the viscosity, 
$\eta = 4\times 10^{-6}\:$Kg/m\,s, of liquid helium, 
the rising bubble has a Reynolds number Re\,$=\rho_f U R/\eta \approx 500$, 
which lies well in the turbulent regime.  
Consequently, when they are released at a depth of about ten 
centimeters, the bubbles need $\approx10$ seconds to rise to the surface.  
This is considerably longer than the time needed for the heat content 
of the bubbles to be carried away.  

We conclude that the helium bubbles condense and release their hydrogen particles 
to the liquid helium before they reach the top of the liquid helium column.  
According to \cite{silvera:1984}, 
the hydrogen particles have no difficulty crossing the liquid-gas interface of helium, 
so that they will move to the liquid phase continuously as the bubbles shrink, 
rather than being caked into a single large aggregate before release.  
Hence, we observe a much larger number of small hydrogen clusters 
than the number of bubbles released into the cryostat.

\begin{figure}
\[
\includegraphics[width=0.45\textwidth]{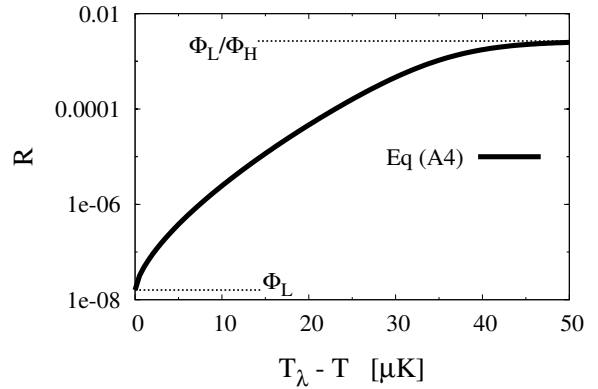}
\]
\caption[] {As a consequence of the rapid rise of the condensation 
  energy $\epsilon$, condensation arises in a very narrow temperature 
  interval. According to our model, \Eq{ratio}, which addresses the
  thermodynamic equilibrium distribution, the vortices will be densely
  packed less than $50\mu$K below $T_\lambda$.
  \label{fig:condensation}}
\end{figure}

\subsection{Thermodynamics of particle trapping}
\label{sec:thermo}

We have already observed experimentally that quantized vortices trap hydrogen particles 
only when the temperature of the helium is sufficiently far below T$_\lambda$.  
We now show that thermodynamics does not explain this link.  
The strength of the interaction between quantized vortices and 
particles is proportional to the mass density of the superfluid part of the helium, 
which rises sharply from zero as temperatures fall below the phase transition temperature.  
However, for some range of temperatures immediately below T$_\lambda$, 
the interaction is still negligible relative to the thermal energy, $k_B$T, 
so that particles reside on the vortex cores only accidentally.  
It follows that as the 
temperature falls, the transition between free particles and 
trapped particles occurs later than the transition from normal to 
superfluid helium.  

In the absence of other effects, such as macroscopic flows, the nature 
of the trapping transition can be determined with a lattice model 
where the lattice constant amounts to the characteristic size of the hydrogen particles.  
In this context, we say that vortices run through a fraction $\phi_V$ of the lattice sites, 
such that there are $N_V = \phi_V N$ sites occupied by vortices, 
and $N_B = (1 - \phi_V) N$ sites making up the fluid bulk, 
where $N$ is the total number of sites in the lattice.  
In the Appendix, we show that the fraction of particles on vortices is then 
\begin{equation}
  R \approx \frac{\phi_V}{\phi_H + \exp(\beta \epsilon)} \, , 
\label{eq:condfrac}
\end{equation}
where $\phi_H$ is the volume fraction of particles, $\beta = 1/k_B$T, 
and $\epsilon$ is the change in the system energy due to the trapping 
of a single particle on a vortex (which is negative).  
The factor $\epsilon$ contains the most important temperature dependence, 
which is that due to the rapid change in density of the superfluid part near T$_\lambda$ 
\cite[e.g.][]{tam:1987}.  
It can be seen in \Fig{condensation} that according to this model, the vortices saturate 
with particles at temperatures only 50\,$\mu$K below T$_\lambda$.  
Because experimentally we observe little particle entrapment up to 20\,mK below T$_\lambda$, 
it follows that the ability to track vortices is not limited by thermodynamic considerations.  
Instead, it is dominated by the dynamics of particle trapping.

\subsection{Dynamics of particle trapping}
\label{sec:dynamics}

At temperatures far enough below the phase-transition temperature, 
quantized vortices begin to gather particles on their cores \cite[][]{parks:1966, bewley:2006}, 
provided the relative velocity between the quantized vortices and the normal fluid 
is not too high \cite[][]{chagovets:2011}.  
We estimate here the time required for a newly-formed length of quantized vortex to collect 
enough particles to make it visible.  

\begin{figure}
\includegraphics[width=3.25in]{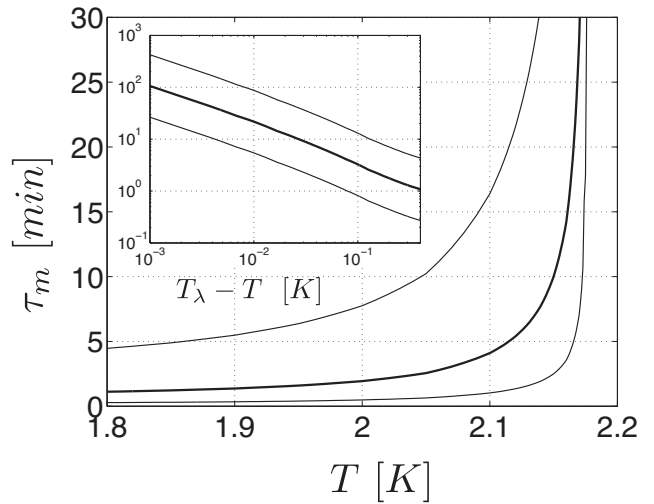}
\caption{ The figure shows the dependence of the vortex marking time,
  $\tau_m$, given by equation~\ref{eq:markingtime} on the temperature.
  Here, the particle diameter, $d$, is 0.5, 1, and 2~microns for the
  lower, middle and upper curves, respectively, and the particle volume
  fraction, $\phi_H$, is 1$\times10^{-5}$.  For lower volume fractions, 
  the marking time is longer.  The inset shows the same 
  curves on a log-log scale, and as a function of the temperature difference, 
  T$_\lambda$ - T.  }
\label{fig:markingtime}
\end{figure}

Consider a stationary vortex that extends through a fluid filled with initially stationary particles, 
whose total volume fraction relative to the fluid is $\phi_H$, and whose characteristic diamter is $d$.  
The distance between particles in the fluid is then typically $l_o = d/(2 \phi_H^{1/3})$.  
At a large distance $r$ from the core, the force exerted by the vortex on the particle is approximately 
      $F_v = -(1/6) \pi d^3 \partial_r p$, 
where $p = -\rho_s \kappa^2 / 8 \pi^2 r^2$ is the pressure, 
      $\rho_s$ is the density of the superfluid, 
      $\kappa = h/m \approx 1 \times 10^{-3}$~cm$^2$/s is the circulation of the vortex, 
      $h$ is Planck's constant, 
and $m$ is the mass of a helium-4 atom.  
Let the vortex be considered marked when all particles within this 
distance from the vortex centerline have reached the core.  
It can be readily confirmed that this corresponds to a core with particles 
every 20 diameters, for a typical particle volume fraction of 10$^{-5}$.  
If the particle's inertia is small, we can balance the 
vortex pressure gradient force by the drag force along its whole trajectory.  
In this picture, the particles are drawn toward the core 
by the superfluid pressure field, but are damped by their motion 
relative to the stationary, viscous normal fluid.  
The drag force for small particles was given by Stokes as $F_S = 3 \pi \mu d v$, 
where $\mu$ is the viscosity of the normal fluid and $v$ is the velocity of 
the particle relative to the normal fluid \cite[][]{batchelor:1967}.  
The velocity, $v$, is induced by the pressure gradient such that 
   $F_v = F_S$.  
Altogether, the time scale for marking is given by integrating the velocity from a 
distance $l_o$ to the vortex core, 
\begin{equation}
  \tau_m = \int_{l_0}^0 \frac{dr}{v} 
         \approx \frac{1}{2} \frac{\mu}{\rho_s} 
         \left( \frac{3 \pi d}{2 \kappa \phi_H^{2/3}} \right)^2.  
\label{eq:markingtime}
\end{equation}
The time scale is temperature dependent through the material 
properties, and particularly through $\rho_s$, which rises from zero 
for temperatures decreasing from the phase transition temperature, T$_\lambda$.  
Figure~\ref{fig:markingtime} shows the dependence of 
$\tau_m$ on temperature for values of $d$ and $\phi_H$ typical in an experiment.  
The sharp decrease in the trapping time as the 
temperature decreases may explain the observation that 
marked vortices typically do not appear within about 20\,mK 
of the phase transition temperature.  
That is, they might become visible if the experimenter waited long enough.  

\section{Discussion and Conclusion}
\label{sec:conc}

The journey of hydrogen from a molecular gas to crystallites that are trapped on quantized vortices 
encompasses a hierarchy of problems on many length scales.  
Novel problems arise 
in applied mathematics 
through the analysis of nonlinear evolution equations, 
in non-equlibrium statistical physics 
through the condensation of hydrogen and the collapse of bubbles, 
and in fluid dynamics 
through the mixing of particles into a fluid, the trapping of particles onto vortices, 
and the turbulent motions of the vortices.  
These challenges call for a combined effort from 
experimentalists and computational physicists to provide data, 
and from theoreticians to identify approaches and questions.  

In this spirit of providing guidance for future experiments, 
where efficient particle generation is desired 
or where manipulation of the particle sizes and densities are necessary, 
we make the following propositions.  
There should be many small injectors, which produce small bubbles 
that disappear quickly.  
The injectors should be as far below the surface of the liquid helium as is possible, 
and the injector tube could be coiled within the cryostat 
in order pre-cool the mixture it flows into the helium liquid.  
The cryostat should be as tall as possible.  
The particles should be as small as possible in order to become trapped quickly.

\acknowledgments

The authors appreciate suggestions from P.~Toennies.  
We also acknowledge discussions with B.~Altaner and J.-H.~Tr\"{o}semeier 
about the thermodynamics of particle entrapment.  
GB gratefully thanks the \emph{Turbulent Mixing and Beyond} committee members, 
in particular S.~Abarzhi and K.R.~Sreenivasan.

%
%


\appendix
\section{Condensation of Hydrogen crystallites on vortices} 
\label{sec:appendix}

To support our discussion of the trapping of the crystallites on the vortices in \Sec{thermo}, 
we estimate their equilibrium distribution based on a simple lattice model, 
where the trapping amounts to condensation into bound states of particles on the vortex.  
The binding energy is $\epsilon$, 
and the particles have a monodisperse size distribution with diameter $d$.  
We assume that the particles do not interact (in particular, they do not merge).

\subsection{Model}

We consider a lattice model with a lattice constant of the order of the particle diameter $d$.  
In a volume $V$ there are 
\[
   N = \frac{V}{d^3}
\]
lattice sites.  
Vortices run through a fraction $\phi_v$ of these 
sites such that there are $N_v = \phi_v N$ sites occupied by a vortex 
and $N_B =(1-\phi_v) N$ sites making up the fluid bulk.  

We choose a description in terms of a grand-canonical potential 
$\Omega(\mu,$T$)$ where T is the temperature and the chemical 
potential $\mu$ will be adjusted to achieve that a fraction $\phi_H$ 
of the sites are occupied by particles. Finally, particles that 
condense onto a vortex change the system energy by an amount $\epsilon < 0$.

\subsection{Partition functions}

For non-interacting particles the grand-canonical potential $\Omega(\mu,$T$)$
is the sum of the functions $\Omega_B$ and $\Omega_V$ describing
particles in the bulk and on the vortices, respectively,
\[
\Omega(\mu,T) = \Omega_B(\mu,T) + \Omega_V(\mu,T) \, ,
\]
They take the respective forms \cite[][]{Callen1985, Widom2002}
\begin{widetext}
\begin{subequations}
\begin{eqnarray*}
  \Omega_B(\mu,T) &=& -k_B T \; \ln \left[ \sum_{n=0}^{N_B}
    {N_B \choose n} \;
    \rme^{ -\beta \mu n }
    \right]
    = -k_B T (1 - \phi_v) \frac{V}{d^3} \; \ln \left[ 1 + \rme^{ -\beta \mu } \right]
\\[2mm]
  \Omega_V(\mu,T) &=& -k_B T \; \ln \left[ \sum_{n=0}^{N_V}
    {N_V \choose n} \;
    \rme^{ -\beta (\mu+\epsilon) n }
    \right]
    = -k_B T \phi_v \frac{V}{d^3} \; \ln \left[ 1 + \rme^{ -\beta (\mu+\epsilon) } \right]
\end{eqnarray*}
\end{subequations}
\end{widetext}
where $k_B$ is the Boltzmann constant.  

The total number of particles, $N_H$, 
is obtained as the negative derivative of $\Omega (\mu,$T$)$ with respect to $\mu$, 
such that the volume fraction of the particles amounts to 
\begin{subequations}
\begin{eqnarray}
  \phi_H & \equiv & \frac{N_H d^3}{V}
  = - \frac{d³}{V} \; \frac{\partial \Omega (\mu,T)}{\partial \mu}
\nonumber
\\[2mm]
  &=& \frac{\phi_V E M}{ 1+EM } + \frac{ (1-\phi_V) M}{1+M}
\label{eq:phiH}
\end{eqnarray}
where
\begin{eqnarray}
  M &=&  \rme^{ -\beta \mu }
\\[2mm]
  E &=&  \rme^{ -\beta \epsilon }
\end{eqnarray}
\end{subequations}

\begin{figure}
\[
\includegraphics[width=0.45\textwidth]{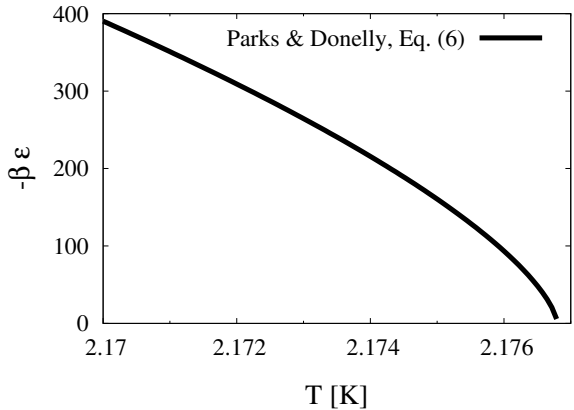}
\]
\caption[]{The condensation energy $-\epsilon$ in units of $k_B$T as a
  function of the temperature (\cf\Eq{epsilon}). 
  \label{fig:energy}}
\end{figure}

\subsection{Fixing the chemical potential}

To compare to the experiments we have to fix the chemical potential, 
\ie the parameter $M$ in \Eq{phiH}, in order to fix the particle 
volume fraction to the particular value $\phi_H$. The exact expression 
for $M(\phi_H)$ can be determined straight-forwardly by multiplying 
\Eq{phiH} by $(1+ME)(1+M)$ and solving the resulting quadratic equation.  

For the following discussion, however, we
observe that $0 \ll \phi_V \ll  \phi_H \ll 1$ in order to find a more
transparent expression. 
Rewriting \Eq{phiH} in the form 
\[
\phi_H  =  \frac{M}{1+M} + \phi_V M \left[ \frac{E}{1+EM} - \frac{1}{1+M} \right]
\]
we see that the expression in square brackets is of order $\beta \epsilon / (1+M)^2$ 
for small $| \beta \epsilon |$ where $E\approx 1$.  
For $\phi_V \ll 1$ one must then
have a value of $M$ of the order of $\phi_H$ and the second term in
the sum may be neglected.  
Consequently, 
\[
M \approx \frac{\phi_H}{1-\phi_H} \; .  
\]
On the other hand, when $\beta \epsilon$ becomes much smaller than
$-1$ such that $E \gg 1$, the second term in the square bracket may be neglected 
and the one in the denominator of its first term will approach $E M$, such that 
\[
M = \frac{\phi_H-\phi_V}{1-\phi_H-\phi_V} \; .
\]
The exact solution varies monotonously between these limits, and in
view of $0 \ll \phi_V \ll \phi_H \ll 1$ we may safely assume in the
following that
\begin{equation}
  M = \phi_H \; .
\label{eq:MM}
\end{equation}

\subsection{Fraction of condensed particles}

From \Eq{phiH} we see that the volume fraction of particles condensed 
onto the vortices amounts to 
\[ 
-\frac{d^3}{V} \; \frac{\partial \Omega_V(\mu, T)}{\partial\mu} = \frac{\phi_V EM}{1 + EM}
\]
To find the fraction of particles condensed to the vortices we divide
by the overall volume fraction of particles, $\phi_H$.  Together with
\Eq{MM},
\begin{subequations}
\begin{eqnarray}
R &=& \frac{\phi_V}{\phi_H} \; \frac{ EM }{1 + EM }
\nonumber
\\[2mm]  
& \approx & \frac{ \phi_V }{ \phi_H + \exp( \beta\epsilon ) }
\label{eq:a4a}
\\[2mm]
&\to& \left\{
  \begin{array}{l@{\quad\hbox{for }}l@{\quad}l}
    \phi_V \, ,        & |\beta\epsilon| \ll 1 & \hbox{\small (accidental)}
    \\[3mm]
    \phi_V/\phi_H \, , & \beta\epsilon \ll -1 & \hbox{\small (full)}
  \end{array}
\right.
\end{eqnarray}%
\label{eq:ratio}%
\end{subequations}

This result states that for $|\beta\epsilon| \ll 1$ a fraction 
$\phi_V$ of the particles reside on sites with vortices.  
In this limit the energy change $\epsilon$ for residing on a vortex is small 
as compared to $k_B$T so that particles do not feel the vortices.  
They are randomly distributed, 
and the particles are found on a vortex only by coincidence.  

In the opposite limit $\beta\epsilon \ll -1$ the particles are
strongly attracted to the vortices.  However, since $\phi_H \gg
\phi_V$, not all particles can be placed on a vortex.  Eventually,
when all sites $N_V$ on the vortices are occupied, the fraction of
condensed particles amounts to the number of lattice sites on
vortices, $N_V = N \phi_V$, over the number of particles, $N_H = N \phi_H$, such
that $N_V / N_H = \phi_V / \phi_H$.

\subsection{Comparison to the experiments}

The cross-over from freely-moving particles to condensing particles arises when 
$\phi_H \approx e^{\beta \epsilon}$, i.e., for $\epsilon \approx k_ B$T$ \: \ln \phi_H$.  
In order to pin down the locus of the transition we observe that 
\begin{eqnarray*}
  \phi_H &=& 6\times 10^{-6}
\\ [2mm]
  \phi_V &\approx& \frac{L_V}{V} \; \Rh^2 
  = \frac{1.6\times 10^4}{\textrm{m}^2} \; \left( 10^{-6}\: \textrm{m} \right)^2
  = 1.6 \times 10^{-8}
\end{eqnarray*}

Finally, the binding energy $-\epsilon/k_B$T is estimated as the
amount of kinetic energy of superfluid which would still be circling
around the vortex core in the absence of the droplet
\cite[][]{parks:1966},
\begin{equation}
\epsilon 
\approx  - \pi \, d \, \rho_s \;
           \left(\frac{\hbar}{m_\textrm{He}}\right)^2 \:
           \left[ 1 
                - \left( 1+\frac{a^2}{d^2} \right)^{1/2} \,
                  \mathrm{asinh}\left(\frac{d}{a}\right) 
           \right]
\label{eq:epsilon}
\end{equation}
where $a \approx 10^{-8}\:$cm is the diameter of the vortex core, 
$m_\textrm{He} \approx 6.688\times 10^{-24}\:$g is the mass of helium atoms, 
and $\rho_s$ is the mass density of the superfluid component of
helium which is approximated according to \citet{tam:1987}.  
Due to the rapid rise of the mass density of the 
superfluid phase upon crossing T$_\lambda$ 
this function rises rapidly when T drops below T$_\lambda$ (\Fig{energy}).
Inserting this data into \Eq{a4a} 
provides the prediction for $R$ shown in \Fig{condensation}.


\end{document}